# A comprehensive stochastic computational model of HIV infection from DNA integration to viral burst

Jayodita C. Sanghvi[1], Don Mai[2], Adam P. Arkin[3], David V. Schaffer[3]

1.  Grand Rounds, Inc.
2.  Stanford University, Department of Computer Science
3.  University of California, Berkeley, California Institute for Quantitative Biosciences

## Abstract:

Multiple mechanisms in the HIV lifecycle play a role in its ability to evade therapy and become a chronic, difficult-to-treat infection. Within its major cellular target, the activated T cell, many steps occur between viral entry and viral burst, including reverse transcription of viral RNA, integration of the viral DNA in the host genome, viral transcription, splicing, translation, host and viral regulation, and viral packaging. These steps exploit complex networks of macromolecular interactions that exhibit various forms of stochastic behavior. While each of the steps of HIV infection have been individually studied extensively, the combinatorial contribution of rare events in each of the steps, and how series of these rare events lead to different infection phenotypes, are not well understood. The complexity of these processes render experimental study challenging. Therefore, we have built a comprehensive computational model of this large system, by collating the community's knowledge of the infection process. It is a stochastic model where rates of different events in the system are represented as probabilities of the event occurring in a timestep of the simulation. This model enables an understanding of the noise and variation in the system. The model also facilitates a dissected understanding of each small part of the large complex system, and its impact on the overall system dynamics.

## Introduction:

Over 1.1 million people in the United States and 34.2 million people world-wide are living with Human Immunodeficiency Virus (HIV).[1] Given the gravity and transmissibility of this disease, HIV has been the focus of research in many labs around the world. The efforts to understand its mechanism of infection, develop critically needed therapies, and eradicate the disease have culminated in a fairly large knowledgebase about this virus. However, molecularly, HIV's infection process is very complex and involves many steps. The majority of research done on this field involves a look at one or a few of these steps. We have developed a tool to try to collate the community's understanding of the molecular biology of HIV into one place to facilitate a systems level understanding of infection. This tool, a computational simulation of the molecular interaction between one HIV-1 particle and one broadly defined T cell, includes every step of infection post DNA-integration. This comprehensive scope is necessary because none of the steps function alone, and we require a better way to understand how the steps together impact infection dynamics. The model is particularly poised to look at the molecular variation between different infection events and its impact on the number of viable progeny generated by a single HIV.
The motivation behind a study of variation stems from the fact that the most common HIV treatments, such as highly active antiretroviral therapy (HAART), are only able to eliminate the actively replicating HIV population. Low levels of latent, or inactive, virus exist in patients' cells, and if a patient halts HAART treatment, these latent populations may re-seed viral infection.[2] Patients must thus receive lifelong HAART treatment, resulting in issues of cost, compliance, drug resistance, side effects, and susceptibility to other diseases.[2] In addition, drug-resistant variants of HIV are emerging as another major challenge to current therapeutics.[2] Our comprehensive look at the system first offers the opportunity to discover new druggable targets to vary the points of intervention and address the issue of drug resistance. Second, the look at the molecular mechanisms behind variation may enable the ability to decrease the variability in the system to 1. greatly increase the probability of generating viable, replicative, progeny that existing treatments can eliminate or 2. greatly decrease the probability of replicative progeny and slow the spread of infection.

HAART and many existing treatments primarily target reverse transcriptase, the viral protein that converts viral RNA into DNA, and viral integrase that integrates viral DNA into the host's chromosomes.[3] To focus on understanding additional targets, this simulation starts at the point of viral DNA integration and aims to explore the molecular variation at each step from integration to viral burst. This is accomplished by exploiting the apparent modularity of biology, and assuming that biological processes are interconnected at long time scales while independent at shorter time scales. Each molecular step of the infection pathway is represented as an independent module in the simulation. The modules, including mRNA transcription, Tat (a positive regulator of viral transcription) feedback, alternative mRNA splicing, Rev (an mRNA transport protein) binding to mRNAs, mRNA export to the cytoplasm, protein translation, mRNA an protein degradation, protein localization, Env (a viral envelope protein) processing, and viral packaging are depicted in Figure 1. The modules are all stochastic representations of the underlying biology and incorporate a combination of novel representations and adaptations/modifications of existing models. Modularity also allows the swapping out of sub-parts and the addition of more detail, as this project progresses. The modules are all integrated at a one minute timestep, and the simulation runs for 48 hours of simulated time, the approximate lifetime of a productive HIV inside of a cell before it kills the cell.[4] The rate constants of critical steps in certain modules are much shorter than a minute, and in these cases those individual modules are run 60 times per "timestep" to simulate variation on the level of seconds. The model explicitly represents each virus derived molecule in the system, but host proteins are assumed to be abundant. The stochasticity is achieved by representing individual rates of events as probabilities of the event occurring at a given timestep. It is important to note that unlike a Gillespie algorithm, this approach only randomizes which events will occur in a timestep and not the length of the timestep itself. The overall methodology used in model construction is very similar to that in Karr et al. 2012.[5]

Lastly, please note that the work presented here was completed in 2014 and no subsequent development has been done. This submission is a report of the model and software developed at that time and may provide a useful starting point for others in modeling HIV and its underlying processes. There are currently no plans to update it for peer review.

## Methods:

A full description of the model, framework, database of parameters, and references associated with each module can be found in the Supplementary Materials. All of the code may be found at https://github.com/donasaur/hiv-model. Here, we will summarize the biology in each of the modules.

First, the HIV reverse transcribed DNA integrates into the host chromosome where it may be transcribed to form viral mRNA. In the transcription module we first consider the location on the genome that the viral DNA integrated. We follow a simple model posed by Skupsky *et al.*, 2010, in which the HIV promoter can be in an ON or OFF rate.[6] Skupsky *et al.* measured the gene expression levels, transcription burst size, and transcription burst frequency in 30 clones with different integration sites.

They then computationally fit the values for the promoter on rate, promoter off rate, and basal transcription rate for each clone. We used these derived values for the constants in our simulation. We make the major assumption that these 30 clones represent the variation of these parameters across all integration sites in correct proportion. For each individual simulation, we randomly pick a number 1-30, and that simulation will run using the Skupsky *et al.* constants for the corresponding clone. Upon basal transcription, viral mRNAs begin to be produced. These "full length" mRNAs will be processed in subsequent modules to mRNAs encoding various proteins of the viral repertoire, and eventually generate viral proteins. One of these proteins, Tat, helps enhance the rate of viral transcription.

In the Tat feedback module, the Tat feedback loop is modeled after the one described in Weinberger *et al*, 2005.[7] We model the binding of Tat to host factor pTEFb, pTEFb acetylation, and its effect on the viral transcription rate. The Tat derived transcription rate is fed into the Transcription module.

The alternative splicing module, at certain probabilities, excises different parts of the full length viral mRNA. Each excision requires host machinery to bind at a donor and acceptor site, which excises the length of mRNA between these two sites. The splicing model is an extension of that presented by Kim and Yin in 2005.[8] Up to 3 splices can occur per mRNA before it translocates out of the nucleus. The first splice event occurs between donor site 1 and one of 7 acceptor sites, leading to shortened mRNAs encoding Vif, Vpr, Tat, Env, Env, Env, and Env. In the second splice event, the *vif*, *vpr*, and *env* mRNAs can be spliced into shortened mRNAs encoding the same protein, or the *env* mRNAs may be spliced to *rev* or *nef* forms. In the third splice event, the doubly spliced *vif* or *vpr* mRNA can be additionally shortened to the *tat*, *rev*, or *nef* forms.

Only the transcripts that have been spliced 2 or 3 times are ready for transport to the cytoplasm. The single-spliced and full-length mRNA must wait for Rev molecules to bind to their Rev response elements (this element has been excised from the multi-spliced RNAs, which is why Rev binding is not a requirement for them). The requirement for Rev binding essentially poses a delay on certain transcripts getting to the cytoplasm. The Rev-independent transcripts start translation before the Rev-dependent transcripts.  The Rev binding module uses data from Pond *et al.*, 2009 to simulate the binding of 8 molecules of Rev to the Rev response element on single-spliced and full-length mRNA.[9] Each binding event has its own probability and binding depends on the availability of Rev and competition from other mRNAs. Rev is also dissociable from this complex, making complex formation a longer process. Upon binding 8 Rev molecules (a fittable parameter), the mRNA is ready for transport to the cytoplasm.

The mRNA export module very simply moves multi-spliced mRNA and mRNA with 8 bound Rev molecules into the cytoplasm at a certain probability.

Once in the cytoplasm, the mRNA may be translated into proteins. The translation module transcribes mRNA into proteins based on a Poisson distribution of the frequency of translation multiplied by the abundance of mRNA. The full length transcript normally encodes the Gag molecule. However, there is a 5% chance of a frameshift at the STOP site for Gag, such that the STOP is passed over and the Gag/Pro/Pol is translated.[10] The various spliced forms encode different proteins as described above in the description of the alternative splicing module. Translation proceeds at its maximal rate until high concentrations of Vpr cause a G2/M cell cycle arrest and the suppression of normal (5' cap dependent) translation. Even when the 5'caps are blocked to ribosome entry on the HIV mRNAs, the full-length mRNAs have a second ribosome entry site (Internal ribosome entry site, IRES) in their 5' untranslated region. Therefore, even after G2/M cell cycle arrest, Gag and Gag/Pro/Pol can still be produced at the frequency of IRES translation.[11] Once translated, the different proteins move from the cytoplasm to different compartments where they function. Rev and Tat act in the nucleus, Env moves into the endoplasmic reticulum (ER) where it starts to be glycosylated, and the other proteins move towards the cell membrane for packaging.

The protein localization module takes into account the shuffling of Rev and Tat between the nucleus and the cytoplasm. Rev and Tat are synthesized in the cytoplasm, but function in the nucleus--Rev to aid in mRNA export and Tat to increase the viral transcription rate. This process stochastically handles the shuffling of these proteins based on the localization rate parameters.

The Env processing module simulates an error-prone multi-step process that converts nascent Env to its functional form. Env (gp160) first moves to the ER where host cell oligosaccharytransferases and glucosidases start to glycosylate Env.[12,13] After the addition of some sugar chains, the chaperone Calnexin will start to fold Env such that additional host glucosidases can further modify Env's sugar chains.[14] At this point, Env is transported to the golgi, where host mannosidases and sugar transferases once again modify Env's sugar chains. All of these glycosylation steps are highly erroneous and it is thought that there is only a 50% chance that the Env molecule is successfully glycosylated at this point.[15] Both correctly and incorrectly glycosylated Env molecules then trimerize, reducing the probability of obtaining trimers with three correctly glycosylated Env molecules. The trimers are then cleaved by a host furin protease into the gp120 and gp41 forms, which then complex back with each other.[16] Lastly, the Env trimers are translocated to the membrane where they can be incorporated into budding virons.[17] Each of the steps in this Env processing module is stochastic based on the rate of the step.

Throughout the simulation, proteins and mRNAs in any compartment may be degraded stochastically based on their decay rates in the Degradation module.

Lastly, the Packaging module simulates the movement of viral and host proteins to the cell membrane and the formation of new viral particles. Both Vpr and Gag levels in a cell have been implicated in the switch that transitions the infected cell from a state of "translating" to a state of "packaging." Here, we include effects of both signals. First, a Vpr threshold is reached that drives the cell into a G2/M arrest.[18] This signal is passed to the translation module as described above. Once in G2/M arrest, cytoplasmic monomeric Gag (not membrane bound) can start to bind to full length mRNAs in the cytoplasm. There are four stem-loops (SL) in the viral RNA that Gag can bind to at different binding constants. Once Gag has bound to SL1, SL2, SL3, the RNA has a conformational change, enabling it to bind to another Gag-bound RNA that also has that conformational change.[18] The whole "nucleated" complex of two RNAs and at least 6 Gag molecules translocates to the cell membrane. Since the translocation time is not known, the model uses a rate roughly inferred by the diffusion time-scale. In the meantime, the remaining cytoplasmic Gag may dimerize and/or migrate to the cell membrane.[19] While trimers and larger complexes of Gag may form, monomers and dimers are most prevalent, and are the only forms included in this model.[19] All of the above steps are modeled stochastically based on rates found in the literature and rates derived from diffusion approximations.

Next, the Gag molecules (the primary structural unit of an HIV particle) start to assemble into budding virons. There is a lot of debate on the number of Gag per HIV particle, and we used an average of 2500 Gag molecules per viron.[20] Assembly is initiated at the site of a "nucleated" complex at the membrane. The Gag monomers and dimers may move from the cytoplasm (slow) or the membrane (fast) to the sites of budding virons. We are modeling exponentially growing budding virons, where the rate of growth in a given timestep is dependent on the cytoplasmic Gag concentration, and the size of the budding viron (cooperative binding). It has been found that higher Gag concentrations yield faster growing virons, and therefore in the model, the exponential rate constant is in part dependent on the cell's Gag concentration at the start of viron growth.[21] For each particle, the final Gag mass is determined using a Poisson distribution of the average Gag mass (2500 molecules). Particle formation is complete and it is ready for budding when this Gag mass is achieved. In the meantime, while the viron's Gag structure is being formed, other HIV and host proteins assemble with it. The rates of incorporation of these other proteins (Vif, Gag/Pro/Pol, Vpr, Nef) are based on the final ratios of these proteins to Gag in the viron.[22,23,24]

All of the parameters used in the model can be found in the parameters.csv file included with the full source code. Any of these parameters can be altered via sensitivity analysis to understand its impact, or altered permanently in the parameters.csv file as new information about HIV molecular biology becomes available.

## Results and Discussion:

Figure 2 displays the first ten hours of a typical simulation and depicts the following expected patterns. Growth of viral mRNA and protein is very slow until a threshold level of Tat protein is produced, at which point the rate of transcription quickly increases (Figure 2, grey dashed line). Transcripts are made in their full length form (encodes Gag or Gag/Pol), but soon may be spliced into forms that encode other proteins. While it takes the longest to generate multi-spliced mRNA, this is the first form to translocate into the cytoplasm as it does not require the viral protein, Rev, for transport. Single-spliced and full length mRNAs must wait until a large complex of Rev molecules bind to them, before we see their entry in the cytoplasm. The Nef, Rev, Tat, Vif, and Vpr viral proteins all have a multi-spliced mRNA form, and therefore are the first proteins that are synthesized. Of these, Nef is produced at the highest abundance, and Vif and Vpr at the lowest. As Rev is produced, the full and single-spliced mRNAs compete for Rev and the eventual formation of an 8-molecule Rev complex at their "Rev response element" site. Therefore, once the Rev levels in the cell are high enough that these complexes start to form, the remaining mRNAs may translocate into the cytoplasm and be translated (Figure 2, grey dotted line). Once enough viral proteins are made, in particular Gag, the primary structural unit of a viron, viral progeny start to be made—only a fraction of which are actually viable. Our proxy for defining a "viable" progeny is it having a threshold level of all of its essential proteins. It is currently defined as having 1 viable Env molecule,[15] at least 1.5% of the average Vpr per viron, and 10% of the average levels of all other proteins in the viron.[22,23,24] These are fittable parameters in the model. The low requirement for viable Env molecules stems from the very low successful Env glycosylation rate.[25,26] Overall, the simulation allows for a look at mass being generated, lost, and moved around the system through the context of various different processes in the system. It enables a visual understanding of different delays and dynamics of the system.

The simulation is built to facilitate sensitivity analysis on any parameter in the system. We performed sensitivity analysis on parameters from each module in the system to understand how tuning those knobs would impact the virus's ability to make virons and noise in the system. Figure 3A-C shows the total and viable progeny produced as we vary the basal transcription rate. Decreasing this rate will increase the time that it takes for Tat feedback of viral transcription to start and increasing this rate decrease the time till Tat feedback kicks in. Varying the rate 10-fold in both directions shows that delaying Tat feedback has a big impact on the eventual production of viral progeny while accelerating Tat feedback does not have a big effect. At higher basal transcription rates, the system may start transcribing faster, but other constraints in the system still control the progeny output. Figure 3D-F shows the effect of tuning the promoter ON rate. This parameter should have similar effects on the onset of Tat feedback, but should also tune the noise in the system. We find that the promoter ON rate has a similar effect on the number of progeny produced from one cell, but higher promoter ON rates decrease the variation in total progeny counts. The variation in viable progeny counts is unaffected, and probably generated by processes downstream in the system.

A second example of sensitivity analysis is depicted in Figure 4 and Supplemental Figure 2 where we vary the rates associated with alternative splicing. Each viral mRNA may be spliced in up to three locations, and each splice event is controlled by its own parameter. The different splice forms encode different proteins. Figure 4A-C shows the effects of changing the probability of the first splice event. As the probability increases, the total progeny generated decreases. This is because Gag, a product of the full length mRNA, is the major structural unit of viral progeny. However, generating the extra Gag comes at the expense of other viral proteins, and therefore the resulting progeny may not have a complete viral protein set, and therefore be inviable. Therefore, there is an optimal probability of the first splice in a given minute that generates the maximum number of viable progeny, estimated to be 0.08 in this simulation. The results of varying the probability of the second splice (Figure 4D-F) show that there similarly is an optimal probability for maximal viable progeny. The thirst splice takes vif and vpr mRNA and splices them into tat, rev, or nef mRNA. Vif and Vpr are the two lowest expressed essential proteins in a viron, and therefore, increasing the third splice form greatly decreases the viable progeny count (Supplemental Figure 2). These patterns start to hint at therapeutic targets. For instance, if one were to reduce the ability of Rev binding to viral mRNAs enough that mRNAs stayed in the nucleus long enough to recieve a third splice, the Vif and Vpr concentration in a cell could be lowered enough to significantly decrease the ability to generate viable viral progeny.

Supplemental Figure 3 shows results of sensitivity analysis on parameters in the other modules in our model. This illustrates how you can probe each part of the system to understand its impact on progeny generation. For example, Supplemental Figure 3I shows the effects of changing the average Gag per viron, a much debated parameter. While many reports measure 1250-2500 Gag per viron, others report 5000 or more Gag per viron. [20,27,28] Our results show that the optimal Gag content for progeny generation is 1250-2500, and that the virus takes a huge hit on progeny production when the Gag content per virus increases beyond this point.

Our simulation serves as a valuable tool for the HIV community. First, it enables a more complete understanding of the molecular mechanisms of HIV and a visual portrayal of system dynamics. Second, it facilitates probing the effect of each variable or set of variables on the system's dynamics as a whole. Third, it explores which variables in the system generate a lot or little variation on other parts of the system. Lastly, it puts all of the system dynamics and variation in the context of viable virus production, to enable an understanding of which factors in the system drive or suppress virality. We present his model as a tool to help bring together various separate analyses of this system, and as more quantitative measurements are made on the system, this model can be expanded to become a stronger and stronger predictor of the system's outcomes. Further exploration using this tool will continue to enhance our understanding of how variation propagates through the HIV infection system, and how one might be able to tune variation for therapeutic goals.

# Figures:

**Figure 1: Schematic of simulated modules and integration.** Depiction of the modules included in the model and when each becomes active over time. Arrows indicate the passing of variables between the modules.

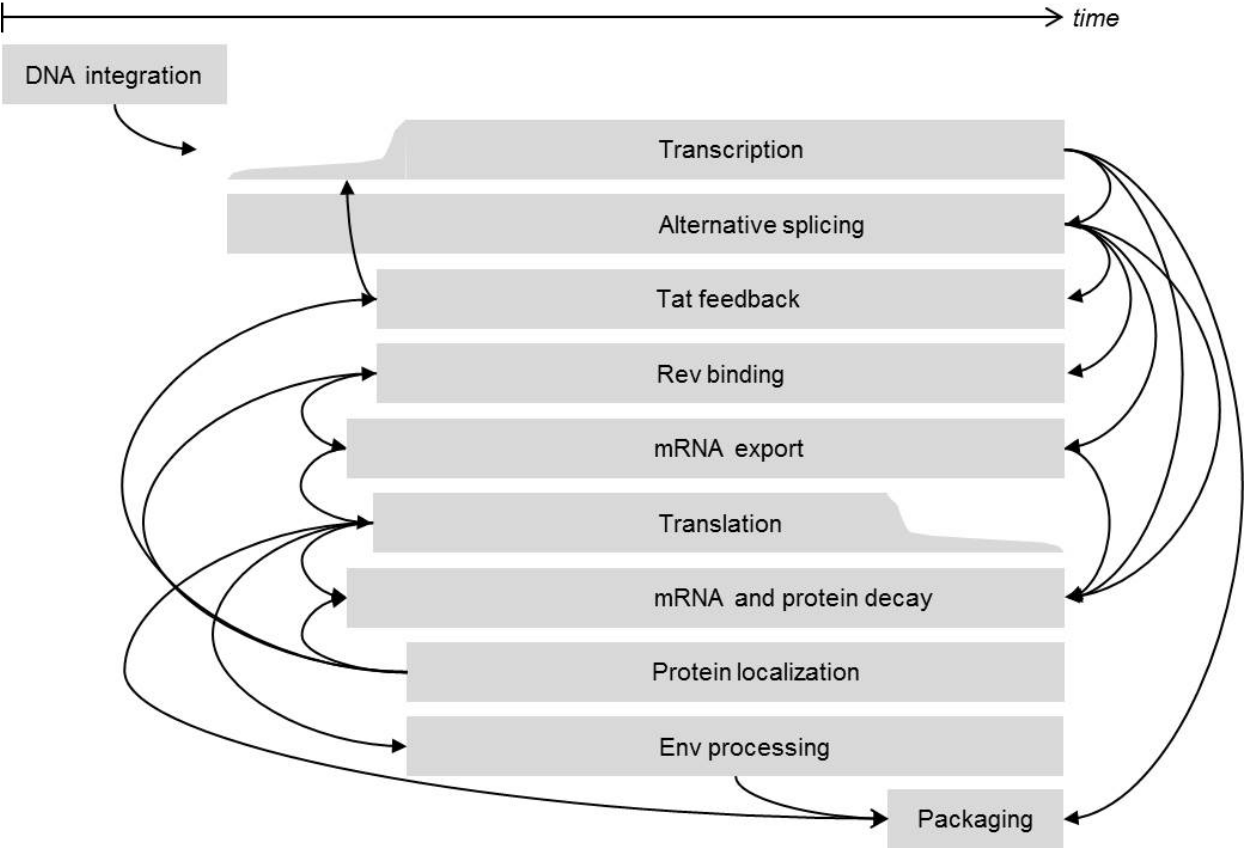

**Figure 2: First ten hours of a typical simulation.** This figure shows the movement of mass through the HIV infection system including mRNA transcription, mRNA splicing, mRNA localization, protein translation, and the formation of viral progeny. The start of Tat feedback is indicated by the grey dashed line, and the start of Rev binding feedback is indicated by the grey dotted line.

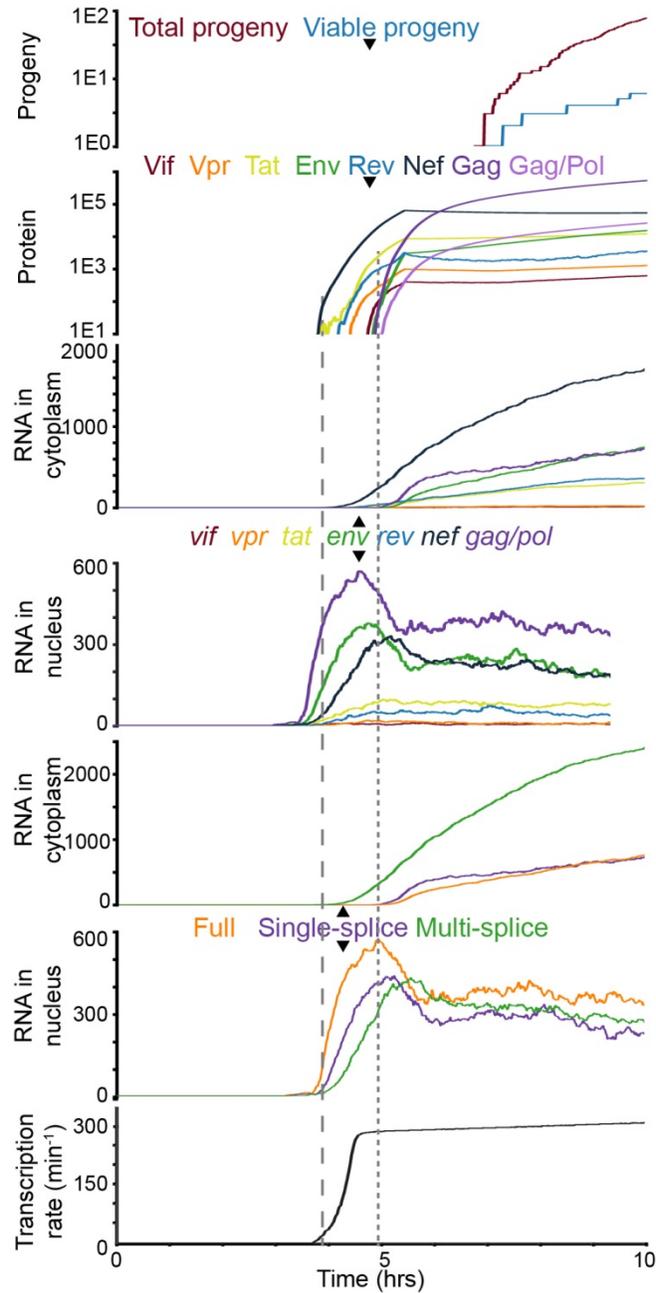

**Figure 3: Sensitivity analysis of basal transcription rate and promoter ON rate.** (A,B) The effect of tuning the basal transcription rate on total and viable progeny production. Five simulations were run for each parameter value and the mean and individual runs are shown. (C) The mean and standard deviation of 5 simulations of progeny produced by the 48[th] hour of simulation. (D-F) Similar plots varying the rate of turning the HIV promoter on.

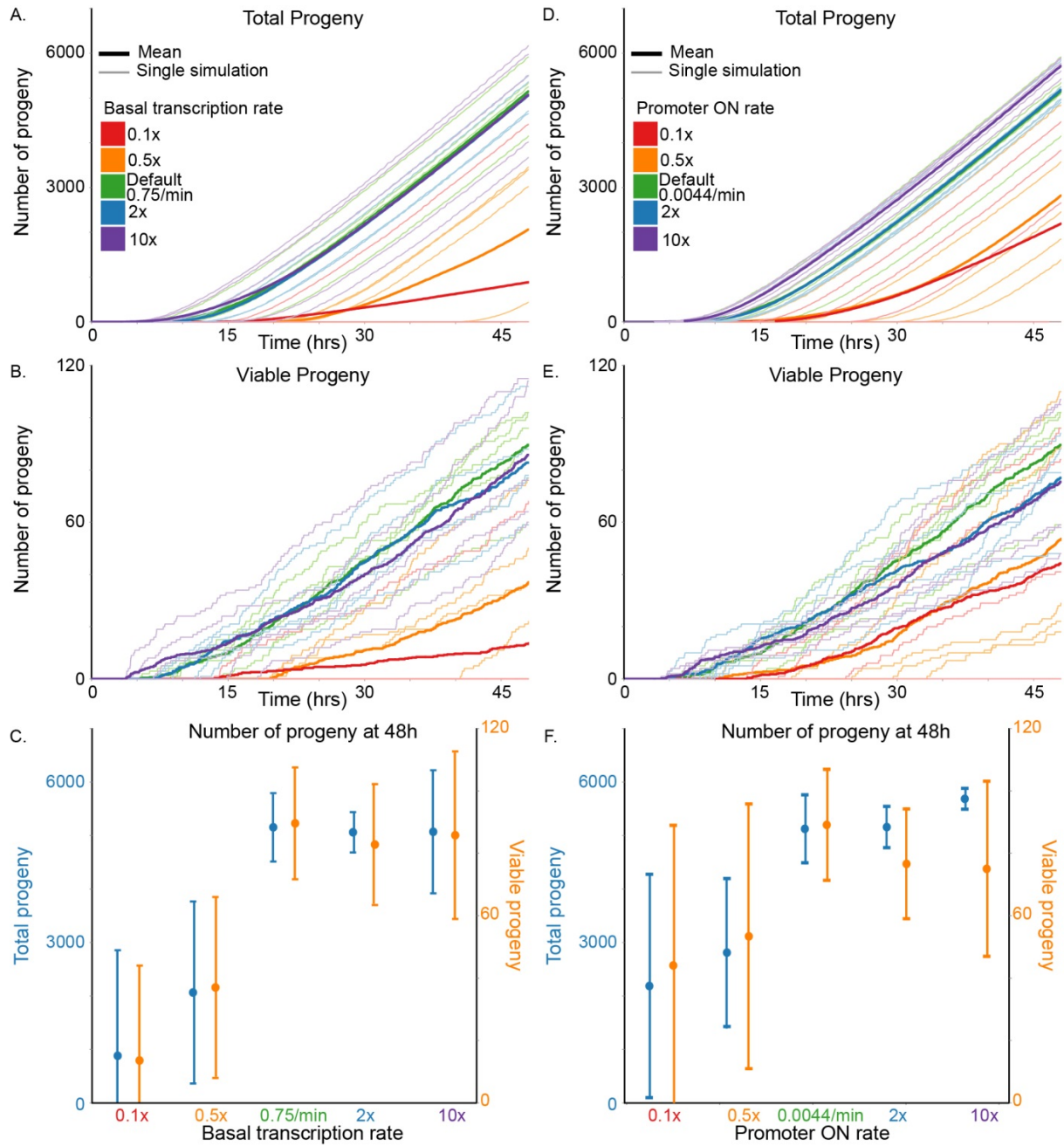

**Figure 4: Sensitivity analysis of alternative splicing probabilities.** (A,B) The effect of tuning the probability of the first splice (from full-length to single-spliced mRNA) on total and viable progeny production. Five simulations were run for each parameter value and the mean and individual runs are shown. (C) The mean and standard deviation of 5 simulations of progeny produced by the 48[th] hour of simulation. (D-F) Similar plots varying the probability of the second splice (single to multi-splice mRNA).

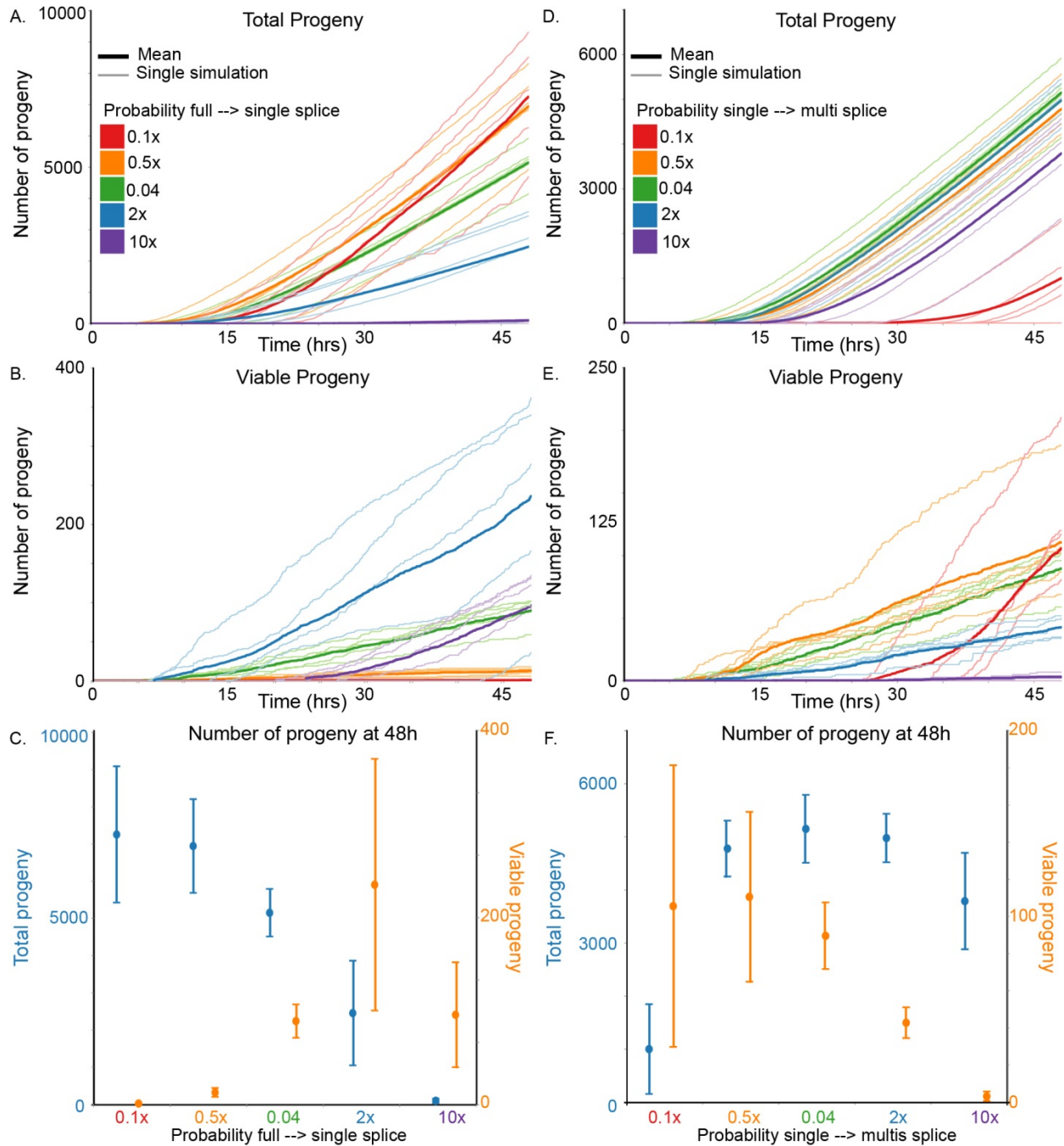

## Supplemental Figures:

**Supplemental Figure 1: Extension of Figure 2 to show an entire 48 hour simulation.**

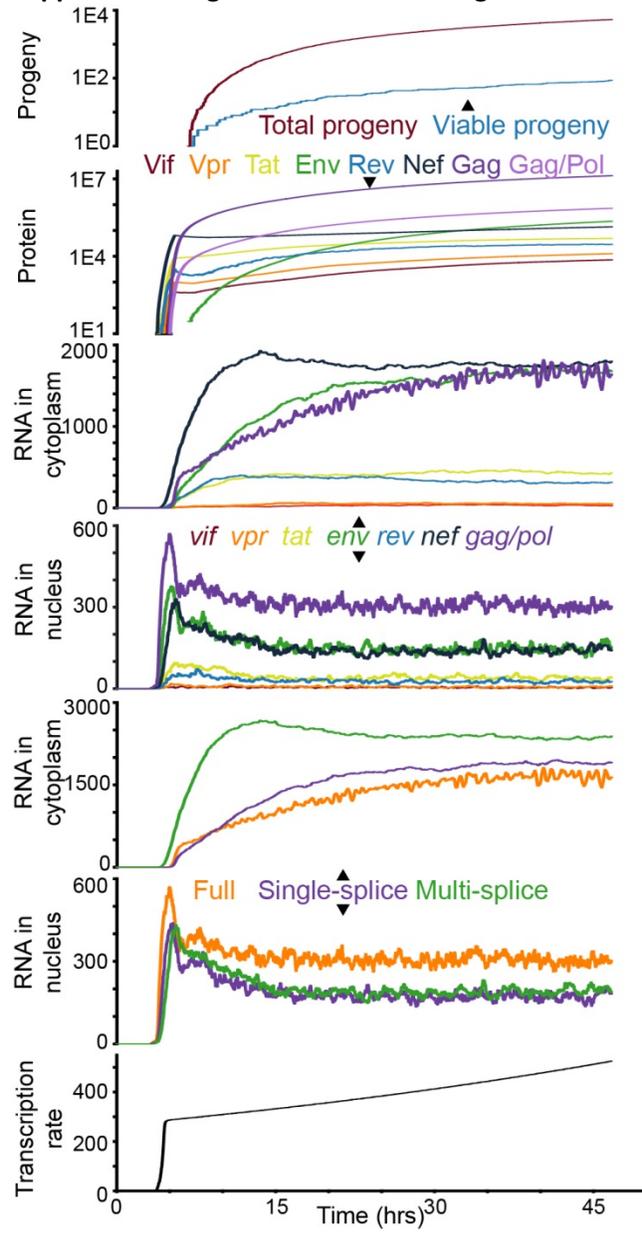

**Supplemental Figure 2: Sensitivity analysis of the third alternative splicing probability.** (A,B) The effect of tuning the probability of the third splice (from double-cut to triple cut multi-spliced mRNA) on total and viable progeny production. Five simulations were run for each parameter value and the mean and individual runs are shown. (C) The mean and standard deviation of 5 simulations of progeny produced by the 48$^{th}$ hour of simulation.

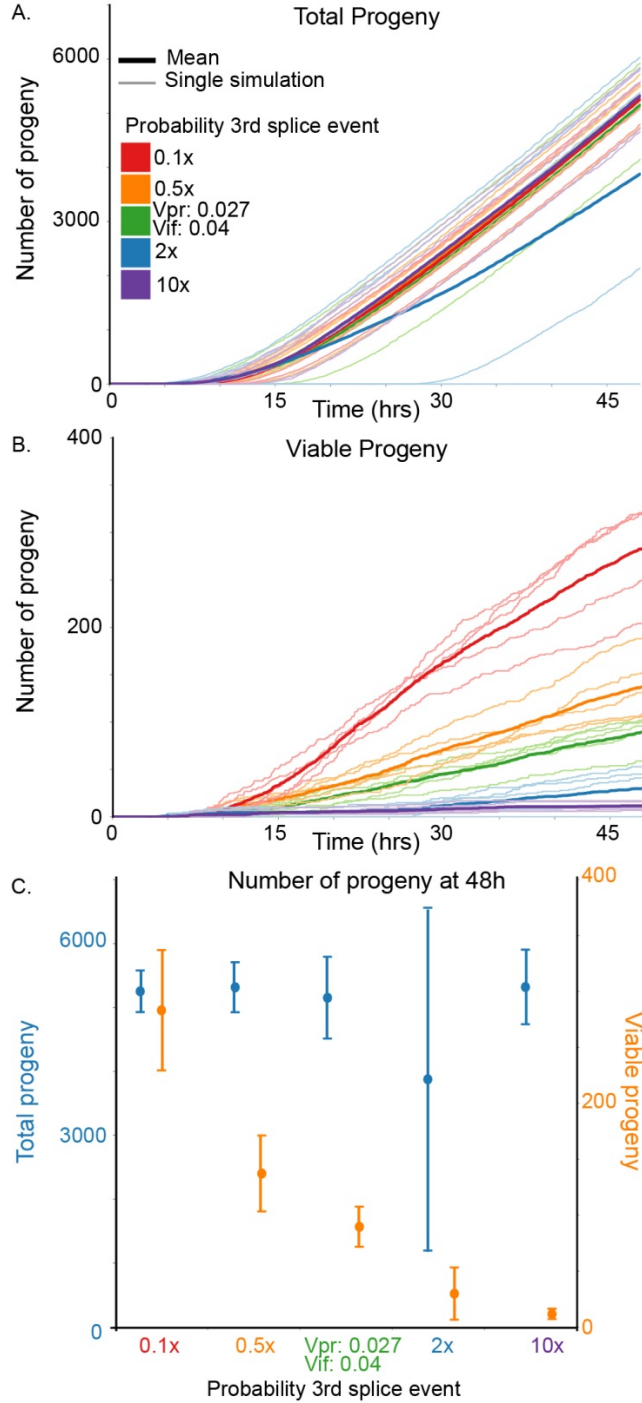

**Supplemental Figure 3: Sensitivity analysis on parameters from various modules.** (A) Tuning the probability of an Env processing error, we see very little effect on progeny output, except when you set the error rate to 100%. This is because the current requirement is set to 1 successfully processed Env molecule per viable progeny. (B,C) Tuning the rate at which Rev and Tat shuttle back into the nucleus from the cytoplasm in the Protein Transport module does not have a large impact on progeny production. (D) The translation frequency as currently set in our model appears optimal for viable progeny generation. (E,F) The mRNA and protein degradation rates have a big impact on viral production, with the mRNA degradation rate having a much larger impact as it has a higher default rate. We tuned the nuclear and cytoplasmic protein degradation rates by the same magnitude in each tuning. (G) Tuning the number of Rev required for full and single-spliced mRNA to translocate to the cytoplasm. While a high Rev requirement (default) generates many progeny, it is interesting to note that no Rev requirement actually generates the same if not more progeny. (H) The Rev binding constant does not play a large role in viral productivity. (I). The optimal Gag per viron is 2500 Gag molecules in the way our model is currently parameterized.

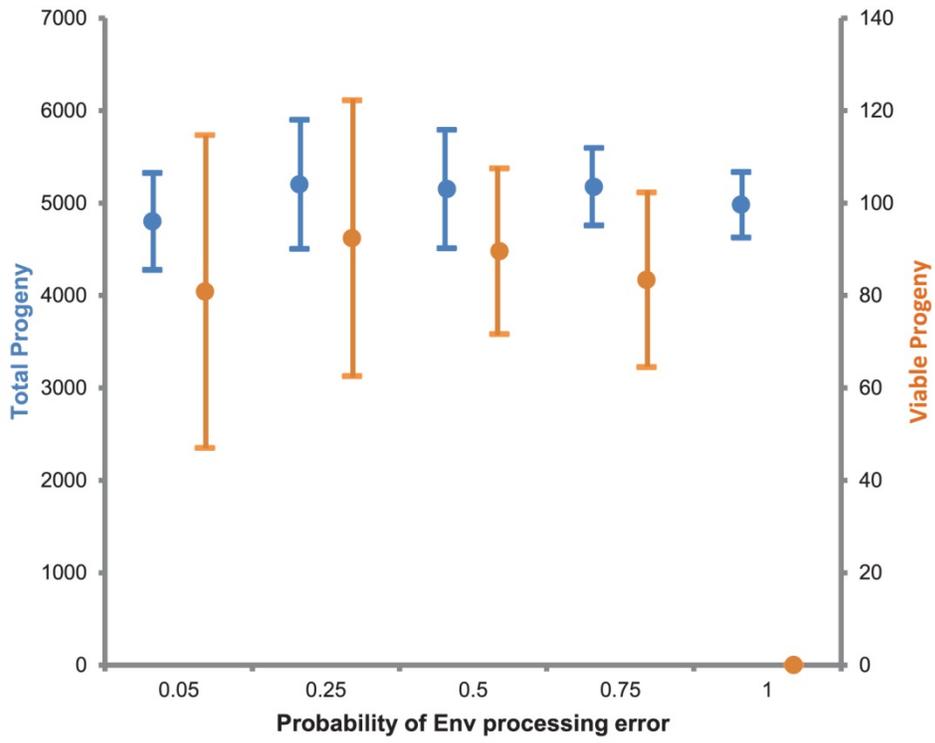
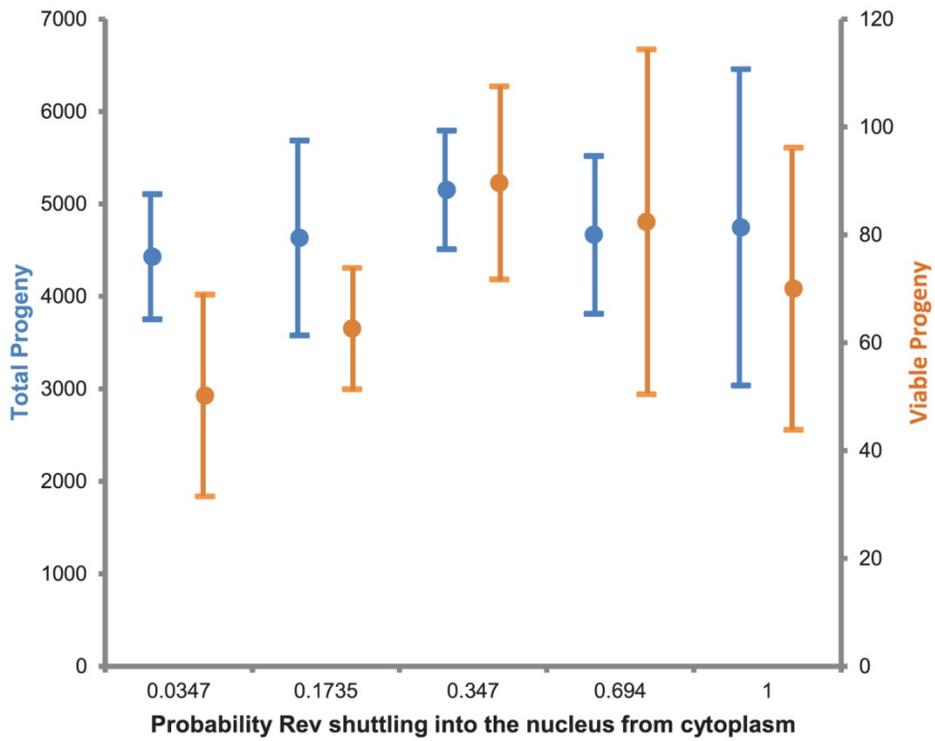

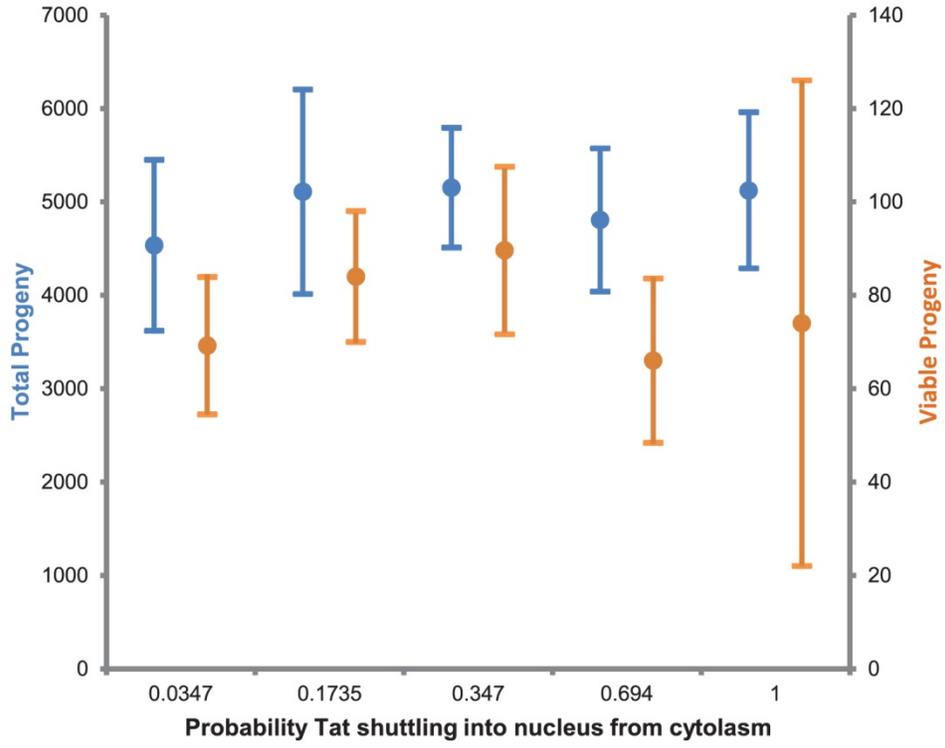

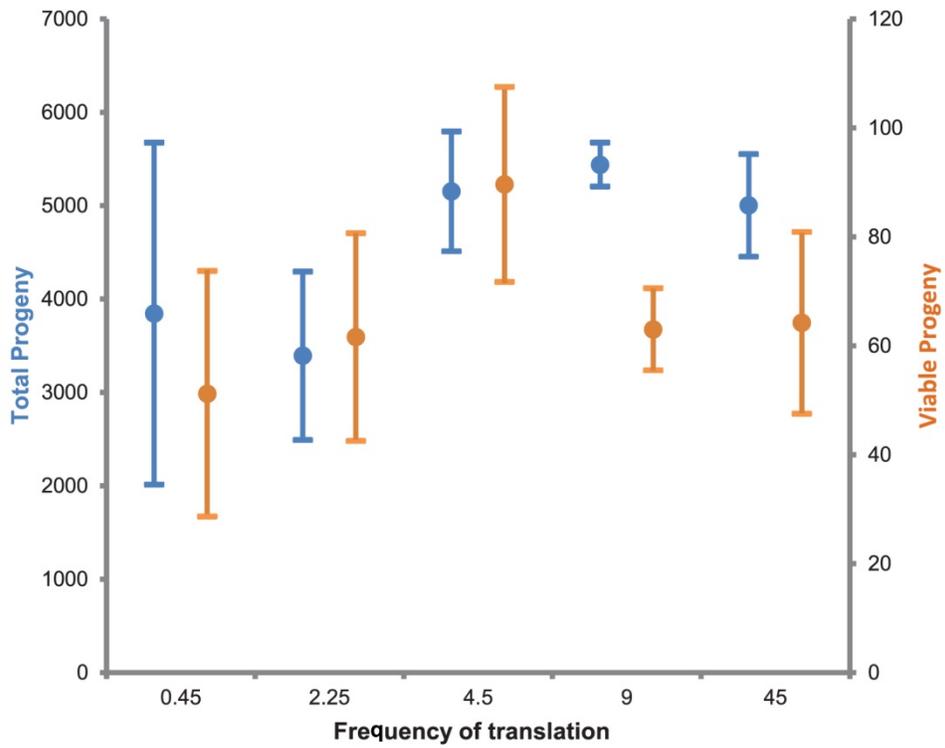

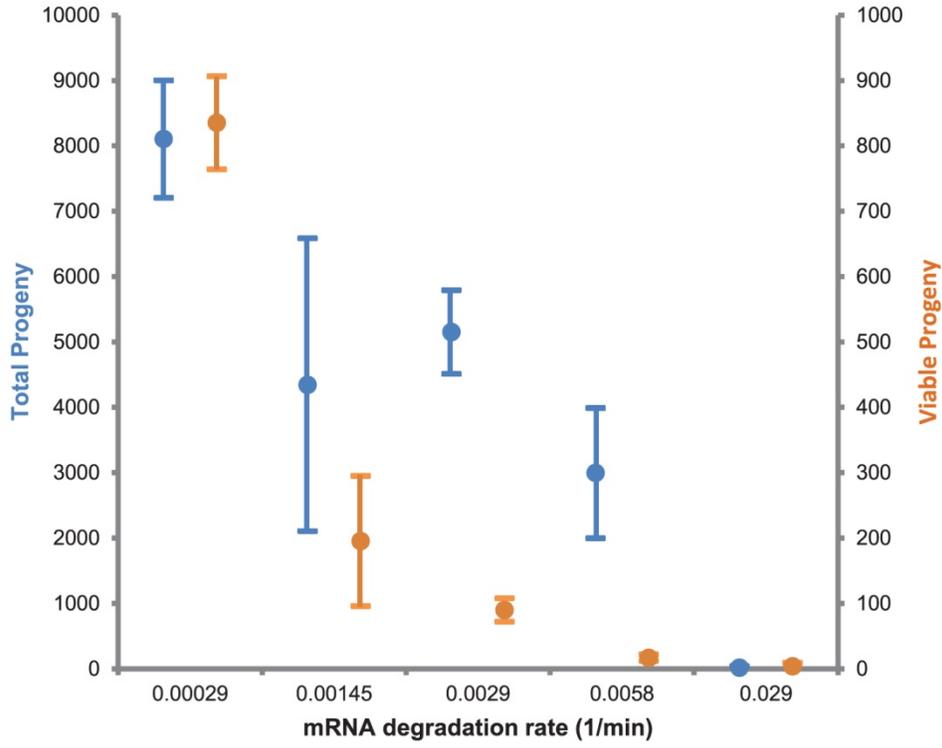

E.

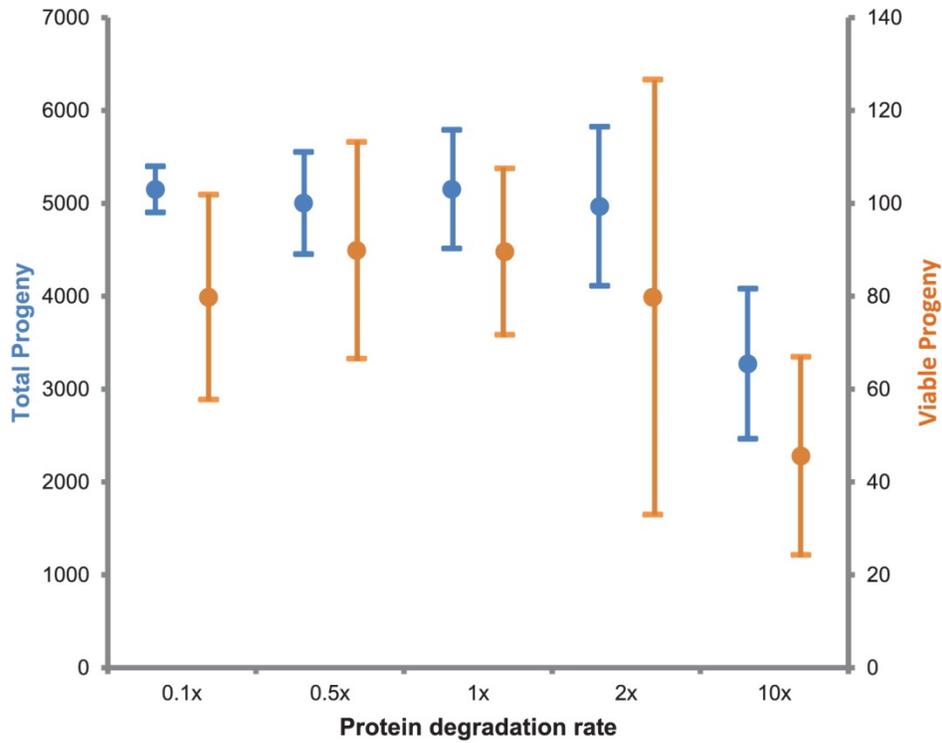

F.

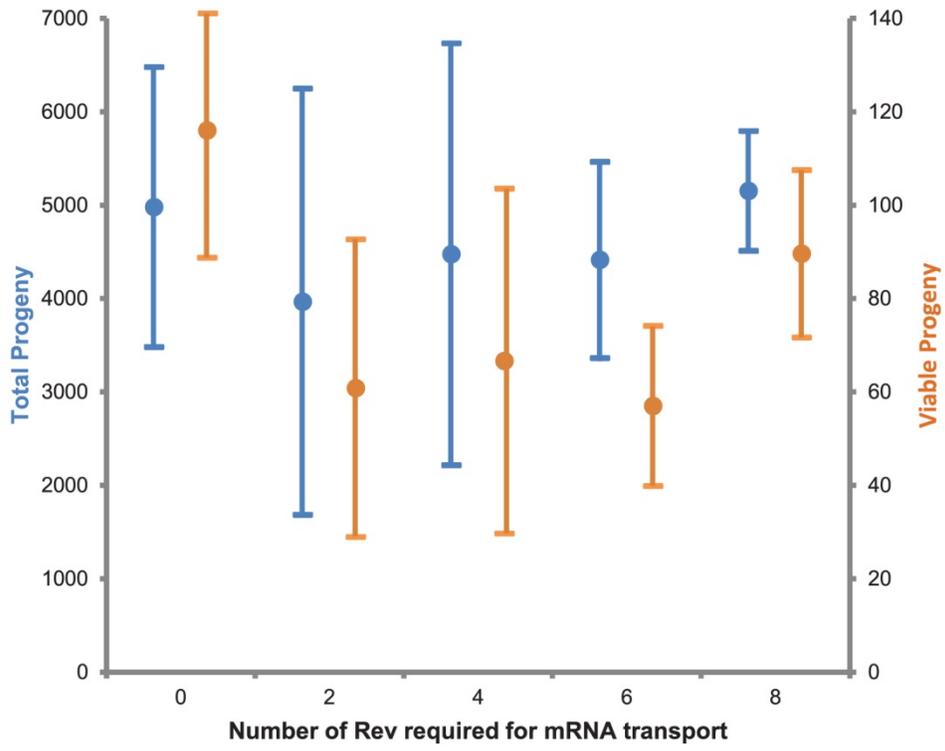

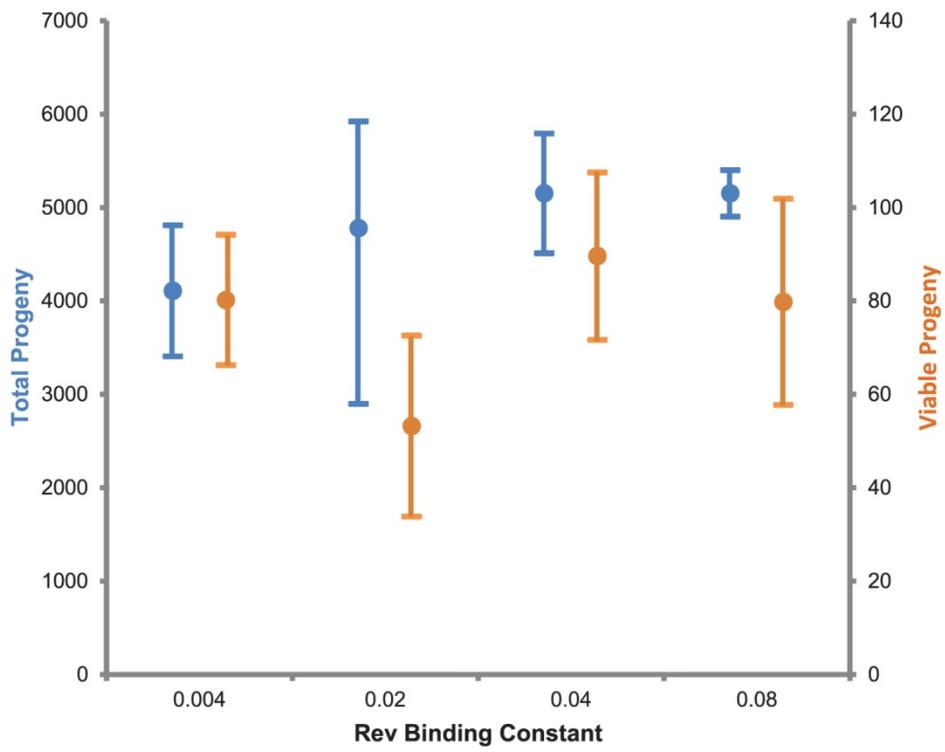

I.

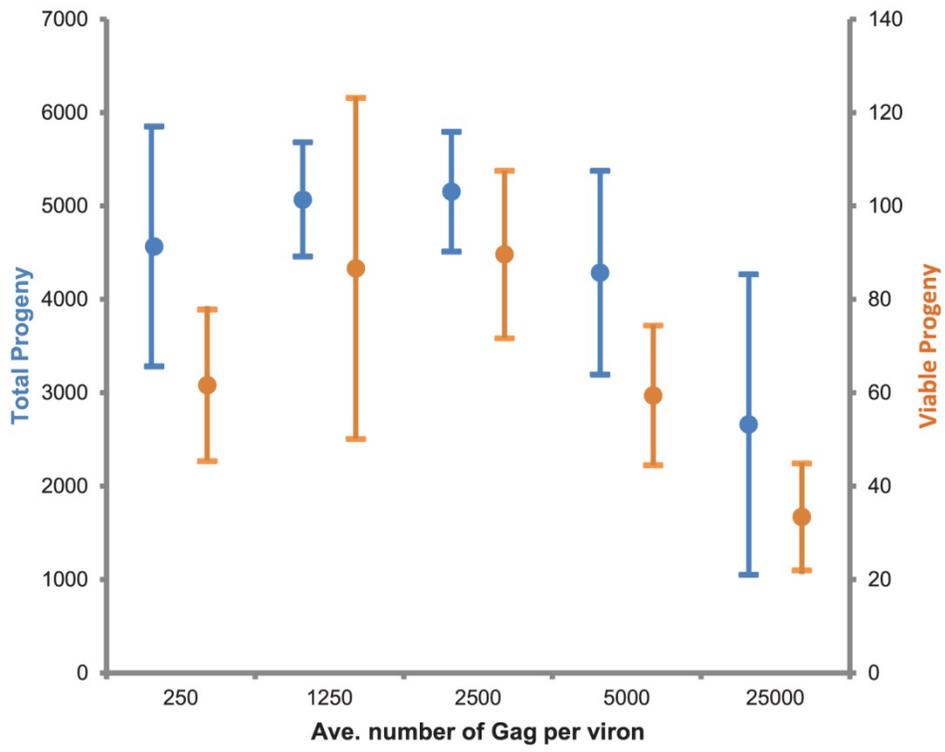

## Supplemental Text:

Please see the docs folder of the project code located at [https://github.com/donasaur/hiv-model](https://github.com/donasaur/hiv-model). The docs folder contains the following documents:

1. Overview – A short overview of the simulator and a listing of features
2. Quick start-up guide – a tutorial on software required to run the model and how to download the model
3. User Manual
    a. Introduction
    b. Plotting guide
    c. Raw data output tutorial
    d. Model structure
    e. Testing
    f. Examples